# Speed and fate diversity tradeoff in nematode's early embryogenesis

Guoye Guan,[1] Ming-Kin Wong,[2] Zhongying Zhao,[2,3] Lei-Han Tang,[3,4,5,*] and Chao Tang[1,6,7,†]

[1] *Center for Quantitative Biology, Peking University, Beijing, China*

[2] *Department of Biology, Hong Kong Baptist University, Hong Kong, China*

[3] *State Key Laboratory of Environmental and Biological Analysis, Hong Kong Baptist University, Hong Kong, China*

[4] *Department of Physics and Institute of Computational and Theoretical Studies, Hong Kong Baptist University, Hong Kong, China*

[5] *Beijing Computational Science Research Center, Beijing, China*

[6] *Peking-Tsinghua Center for Life Sciences, Peking University, Beijing, China*

[7] *School of Physics, Peking University, Beijing, China*

[*] lhtang@hkbu.edu.hk

[†] tangc@pku.edu.cn

*Abstract.* — Nematode species are well-known for their invariant cell lineage pattern during development. Combining knowledge about the fate specification induced by asymmetric division and the anti-correlation between cell cycle length and cell volume in *Caenorhabditis elegans*, we propose a model to simulate lineage initiation by altering cell volume segregation ratio in each division, and quantify the derived pattern's performance in proliferation speed, fate diversity and space robustness. The stereotypic pattern in *C. elegans* embryo is found to be one of the most optimal solutions taking minimum time to achieve the cell number before gastrulation, by programming asymmetric divisions as a strategy.

*Introduction.* — From a fertilized egg to a juvenile, metazoan embryogenesis goes through several distinct stages of development. The maternal-to-zygotic transition (MZT), where maternal gene products are progressively replaced by zygotic ones, separates an initial phase of rapid cell proliferation from gastrulation where morphogenesis begins [1, 2]. Over the long history of evolution, organisms have explored many different schemes of pacing cell proliferation with differentiation to optimize their developmental program [3, 4]. In nematode *C. elegans*, MZT [5] and gastrulation [6] takes place at the 26-cell stage. However, cell fate specification starts already at the first cleavage [7, 8]. The time course of subsequent cleavages are meticulously orchestrated, including reproducible division timing, volume segregation, and migration trajectory of each and every cell [7-10].

Figure 1(a) shows 3D time-lapse images of a wild-type embryo with GFP-marked cell nucleus and mCherry marked cell membrane taken in our study, where details of the experimental procedure is given in Supplemental Material (SM). The somatic lineages AB, EMS and C each proliferate through a sequence of symmetric and synchronized cleavages where daughter cells acquire nearly the same volume and an interim fate identity [Figs. 1(b-c)] [7, 11]. The germline cells P0-P3, on the other hand, divide asymmetrically and generally have longer cell cycles than their somatic siblings [12]. Figure 1(d) presents the cell cycle length against cell volume determined in our experiments. This anti-correlation has been reported in several previous studies of *C. elegans* early embryogenesis [13-15]. While somatic lineages appear to progress at high speed afforded by their larger volume, the smaller germline cells undergo elongated cell cycles that are at least partially attributed to the disparate volume partitioning in asymmetric divisions [Fig. 1(c)]. Thus an asymmetric division, following a polarity cue in the mother cell [16], confers not only different fates to the daughter cells, but also their respective cell cycle lengths.

The synergistic integration of fate diversification and cell division timings through asymmetric cell division is an effective way to bypass genetic instructions to achieve a specific lineage pattern rapidly. In this work, we explore this synergy further by extending the lineage development program of *C. elegans* to a broader class of volume segregation ratios during early embryogenesis. We evaluate each lineage pattern's performance in terms of its proliferation speed, fate diversity and space robustness. The *C. elegans* lineage pattern is shown to be highly optimized, but alternative patterns with similar or even greater proliferation speed also exist, prompting us to carry out further investigation.

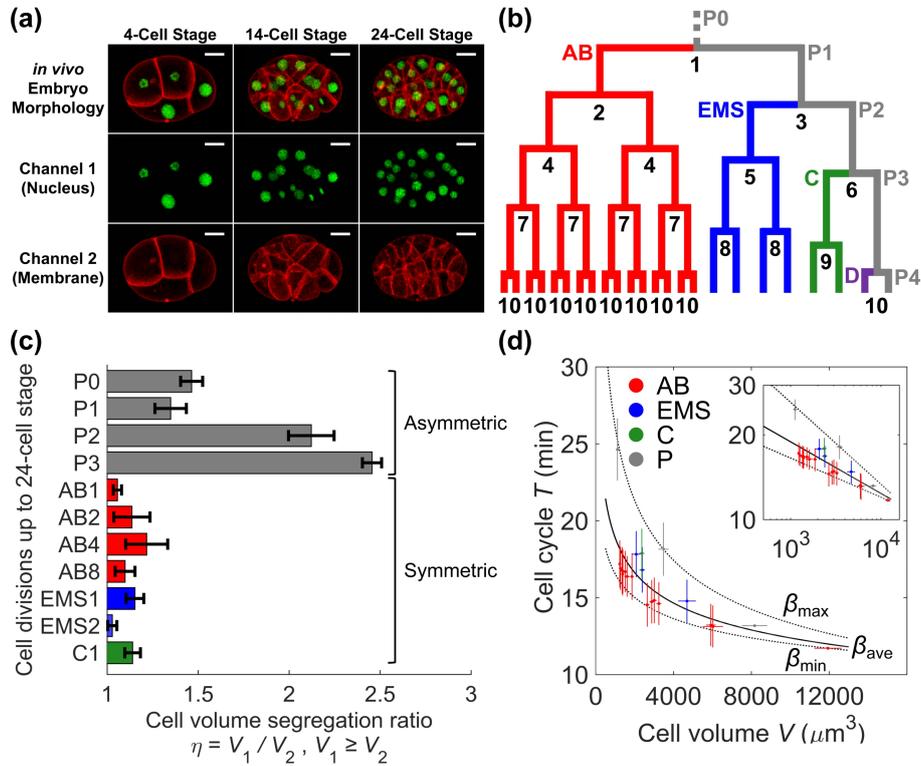

FIG. 1. *C. elegans* embryogenesis up to 24-cell stage. (a) *in vivo C. elegans* embryo morphology illustrated with fluorescence on cell nucleus (GFP) and cell membrane (mCherry); scale bar, 10 μm. (b) Lineage pattern with invariant division order and fate specification. Numbers under each division indicate the order of synchronized cleavages. (c) Cell volume segregation ratio; error bar, standard deviation. (d) Anti-correlation between cell cycle length and cell volume fitted with power-law curves; inset, data shown on logarithmic scale.

*Logistic cleavage model.* — Taking *C. elegans* embryonic development as a reference, we developed a cleavage timing model to explore a much greater space of cell lineage patterns. Each lineage pattern is generated from a set of volume segregation ratios $\eta_n$ at node $n$ of the genealogical tree [Fig. 1(b)], starting from the zygote at the top. The volume $V$ of a descendant cell is obtained from the sequence of cleavages along its ancestral path, with $V_0$ the volume of the embryonic egg. The cell cycle length $T$ is computed using the equation [13],

$$T = T_0 \, (V/V_0)^{-\beta}. \tag{1}$$

From Fig. 1(d), we see that the exponent $\beta$ has a weak lineage dependence, but generally lies in the range 0.14 to 0.29. For simplicity, we adopt a common value $\beta = 0.18$ in this work. The volume and time scales, although not important for our theoretical study, take values $V_0 = 2 \times 10^4$ μm$^3$ and $T_0 = 10.9$ min, respectively, based on our *in vivo* experiments on *C. elegans* [9, 10]. While the biophysical origin of the scaling relation (1) is not yet fully understood [13-15, 17], a relatively weak yet significant dependence of $T$ on $V$ (i.e., a small $\beta$ value) turns out to be what is needed to generate a rich and viable set of lineage patterns. For a given set of volume segregation ratios $\{\eta_n\}$ along the genealogical tree, the above rules yield birth and division timings of all cells on the tree. The elapsed time to reach $N$ cells coincides with the $(N-1)$-th division $t_{N-1}$. Hence the proliferation speed or rapidity is given by $P = 1/t_{N-1}$.

*Fate specification.* — Cell fate determination is a complex process that mostly follows the lineage path but could be affected by other factors such as the local cellular environment. Here, we focus on the fate diversification during the early embryogenesis and specify the number of distinct fates $F$ in an $N$-cell embryo solely based on the distribution of asymmetric divisions in the lineage pattern. Starting from $F = 1$ for the zygote, each symmetric division propagates a given lineage to the next generation without diversification. On the other hand, an asymmetric division creates two new lineages. When the mother cell is the only carrier of the previous lineage, $F$ increases by one. Otherwise, $F$ increases by two. An exception to the last case is when two or more sister cells or cousins in the same lineage divide asymmetrically at the same time, in which case daughters of larger volume share the same fate and the smaller ones another fate. This scheme of fate specification is further explained in SM.

By analyzing images illustrated in Fig. 1(a) from different embryos, we found that symmetric division in the somatic cell lineages may also yield daughter cells of slightly different size, attributable to random fluctuations. Therefore we introduce a division asymmetry threshold $\eta_c$ such that two daughter cells are considered as founder cells of new lineages only when the volume segregation ratio $\eta$ of the mother exceeds $\eta_c$. Based on the data in Fig. 1(c), we set $\eta_c = 1.28$ to separate symmetric and asymmetric cleavages [9].

*Space robustness.* — Continuous addition of new cells following sequential cleavages could jeopardize the canonical cell movement during mechanical equilibration inside the egg, leading to defect patterns [14, 18]. Experimentally and in model studies, synchronization of cell divisions was shown to rescue robust spatial organization. To enforce this property among lineage patterns generated from our model, we introduce two time constants: $\delta_s$ for cleavage clustering and $\delta_a$ for minimal temporal separation between clusters. The parameter $\delta_s$ needs to be sufficiently small so that new cells created within this time interval will move coherently to their equilibrium positions. In contrast, $\delta_a$ is chosen to be longer than the typical completion time of the equilibration process. Based on our experimental cell trajectory analysis [10], we set $\delta_s = 1.5$ min and $\delta_a = 3.0$ min.

Figure 2 illustrates clustering of cell division events along the time axis as required by the above rule. Each cluster corresponds to a burst of new cells born synchronously. The temporal pattern can be considered as a projection of the lineage pattern [e.g., Fig. 1(b)] onto the time axis. In the following, we shall use cleavage clustering to classify and compare lineage patterns that satisfy the space robustness condition.

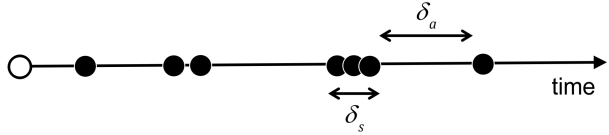

FIG. 2. A sequence of cell cleavage events during early embryogenesis whose timings are illustrated by solid circles along the time axis. Events that fall within a window of size $\delta_s$ form a cluster while neighboring clusters or isolated events need to be separated by a minimal time interval $\delta_a > \delta_s$ to ensure punctuated cell movement.

*Model exploration.* — We devised a two-round sampling scheme to explore the space of lineage patterns, focusing on the 24-cell embryo near the end of the MZT. To give sufficient weight to symmetric divisions, a truncated gaussian distribution is adopted for the volume ratio $\eta = \max(1, \xi)$, where $\xi$ is a gaussian random variable with mean $\mu$ and standard deviation $\sigma$.

The detailed simulation protocol is presented in SM, Fig. S2. In the first round, we take $\mu = \sigma = r\eta_{max}$, where $\eta_{max} = 3.21$ is the maximum cell volume segregation ratio detected among the first seven generations of cells in the experiment [9]. Here $r$ is fraction taking values from 1/6 to 1 in steps of 1/6. For each $r$, a total of $Q = 5 \times 10^5$ independent sets of $\{\eta_n\}$'s are generated and then filtered under the clustering criterion shown in Fig. 2. In the second round, we sample the space around the first round solutions $\{\eta_n\}$ by substituting $\eta_n \rightarrow \eta_n' = \max(1, \eta_n + \xi')$, where $\xi'$ is another gaussian random variable with mean 0 and standard deviation $r\eta_{max}$. The procedure is repeated $Q$ times for each solution from the first round. Results for different $r$ values are then merged together to yield a final set of 19,654 viable lineage patterns.

Figure 3(a) presents our lineage solutions in the *P-F* plane, which exhibit considerable spread. The *C. elegans* lineage pattern shown in Fig. 3(b) is located in the lower right corner of the plot (Pattern No. 2 indicated by asterisk). Its proliferation speed ranked among the top 0.2%. The remarkable performance persists when Eq. (1) is replaced by other functional forms, supporting the general validity of this conclusion [SM, Table S2].

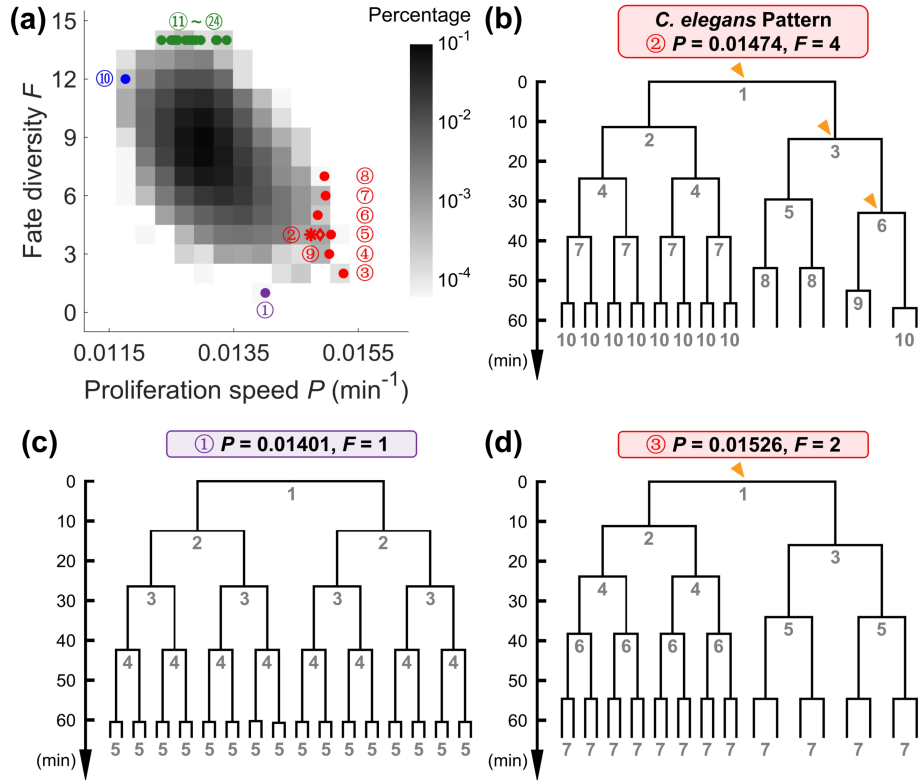

FIG. 3. Lineage solutions at $N = 24$ from random sampling. (a) Distribution of the 19,654 non-identical solutions in $P$-$F$ space; lineage patterns with the lowest and highest fate diversity (Patterns No. 1 and Nos. 11-24), the lowest and highest proliferation speed (Patterns No. 10 and Nos. 3-8) are indicated with purple, green, blue and red points, respectively; the *C. elegans* pattern is denoted by red asterisk (Pattern No. 2). A complete list of the labeled lineage patterns can be found in SM, Figs. S3 and S4. (b) The *C. elegans* pattern. (c) The fully symmetric solution (Pattern No. 1). (d) The fastest solution (Pattern No. 3). Orange triangles denote the asymmetric divisions with $\eta > \eta_c$.

Fate diversity can be tuned extensively in our model. The lineage pattern with the least diversity ($F = 1$) consists of only symmetric divisions [Pattern No. 1 in Fig. 3(c)]. However, some can accommodate 14 fate specifications by introducing 11 to 13 asymmetric divisions during development [green dots in Fig. 3(a) and SM, Fig. S4]. Also seen is a tiny fraction of solutions that proliferate faster than the *C. elegans* pattern [red dots in Fig. 3(a)]. An example is shown in Fig. 3(d) (Pattern No. 3). Their fate diversities are limited to the range 2 to 7 (see SM, Fig. S3). Overall, the data exhibits a statistical tradeoff between the proliferation speed $P$ and fate diversity $F$.

*Speed optimization.* — To better understand the oval-shaped distribution seen in Fig. 3(a), we start with Pattern No. 3 at the lower right corner which has only two lineages following the first cleavage. Timing of subsequent cell proliferation events can be easily computed in our model. Figure 4(a) shows our results for selected values of the zygotic volume segregation ratio $\eta_1$. Two sample lineages with smaller and larger values of $\eta_1$ are shown as inset. At specific time points, the number of cells in the symmetric case ($\eta_1 = 1$) changes from $n$ to $2n$, where $n$ is a power of 2. Asymmetric division creates a larger and a smaller blastomere which become founder cells of respective lineages. In each generation, $n / 2$ cells have slightly greater volume than the other half and hence divide at an earlier time. Consequently, the embryo reaches $3n / 2$ cells faster than the fully symmetric case. The latter series includes the 24-cell embryo. Tuning up $\eta_1$ shortens the time for the faster lineage to reach 16 cells, until it overlaps with the previous event where the slower lineage turns into 8 cells. At this point, we arrive at Pattern No. 3 that leads all other lineage patterns in speed and grows approximately 6 min faster than the fully symmetric pattern [Fig. 3(d)].

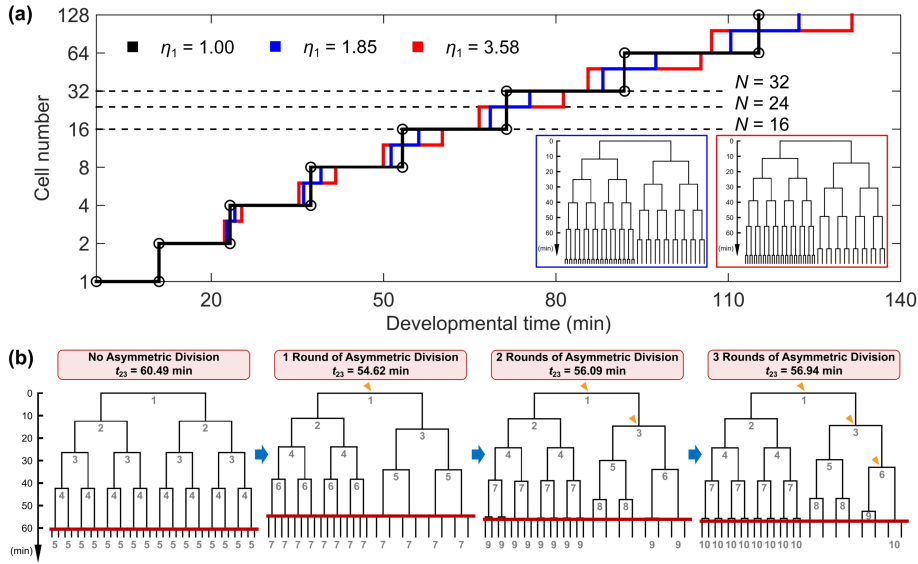

FIG. 4. Impact of asymmetric divisions on proliferation speed. (a) The curves of cell number over developmental time in the lineage patterns with only asymmetry on the initial division. (b) The evolution from the fully symmetric pattern to the *C. elegans* pattern; red lines denote the timing reaching 24 cells; orange triangles denote the asymmetric divisions with $\eta > \eta_c$.

Likewise, the *C. elegans* pattern can be considered to have evolved from the fully symmetric case by adding three asymmetric divisions onto its smaller blastomere sequentially [Fig. 4(b)]. In theory, any additional asymmetric division prolongs the time for the smaller blastomere to reach 8 cells, thus decreasing the proliferation speed. Nevertheless, the total time loss between Pattern No. 3 and the *C. elegans* pattern is less than 2.5 min. It is worth pointing out that, when the target cell number is a power of 2, the fully symmetric pattern always has the highest *P* value. These findings together explain the counterintuitive phenomenon that meticulously programmed asymmetric divisions make the embryo grow faster than the fully symmetric one, while the *P-F* tradeoff is maintained statistically.

We have checked that the above scheme applies to all fast-proliferating lineage patterns at the lower right corner of Fig. 3(a) (see also SM, Fig. S3). Their zygote divides asymmetrically and generate two fate-specific blastomeres; then the larger one undergoes symmetric and synchronous divisions, and produces identical descendants rapidly, while the asymmetric divisions occur in another blastomere (Patterns Nos. 2-5 and 9). Alternatively, the asymmetric divisions can be programmed in the larger blastomere as well to allow more cell fates while the proliferation speed is only slightly reduced (Patterns Nos. 6-8).

*Lineage interference.* — The fast-proliferating solutions have a relatively small number of lineages. Within each lineage, cell divisions are synchronized, hence only a few temporal clusters of cleavage events are needed to reach the target cell number $N$. On the other hand, for a solution with many lineages or equivalently a large $F$, subsequent cell divisions are much harder to coordinate and possibly in conflict with one another under the minimal separation requirement of Fig. 2. We name this phenomenon "lineage interference".

Figure 5(a) shows the number of distinct lineage patterns $\Omega(N)$ found in our model simulation. The curve grows exponentially at small $N$ but decays dramatically after reaching its peak at $N = 24$ to 25. The mean fate diversity $F(N)$, averaged over the solutions, also show a change of behavior around $N = 24$ [Fig. 5(b)]. To examine this crossover in more detail, we introduce two differential quantities, the inheritance rate $r_I$ and expansion rate $r_E$. At a given $N$, $r_I(N)$ represents the percentage of patterns that can be further grown to $N + 1$, while $r_E(N)$ represents the ratio of patterns at $N + 1$ to their original and inheritable patterns at $N$. As shown in Figs. 5(c)-(d), the two quantities have a similar dependence on $N$. Before reaching the 16-cell stage, the majority of patterns can be extended further ($r_I \approx 0.99$). The expansion rate $r_E$ exhibits an even-odd oscillation, with an average value around 1.76. From $N = 16$ onward, both $r_I$ and $r_E$ start to decline. At $N = 23$, $r_I(N) \times r_E(N) \approx 1$ that marks the peak of $\Omega(N)$. Interestingly, the peak location coincides with near completion of the MZT in *C. elegans*, beyond which Eq. (1) is subject to lineage-specific modifications [5, 15].

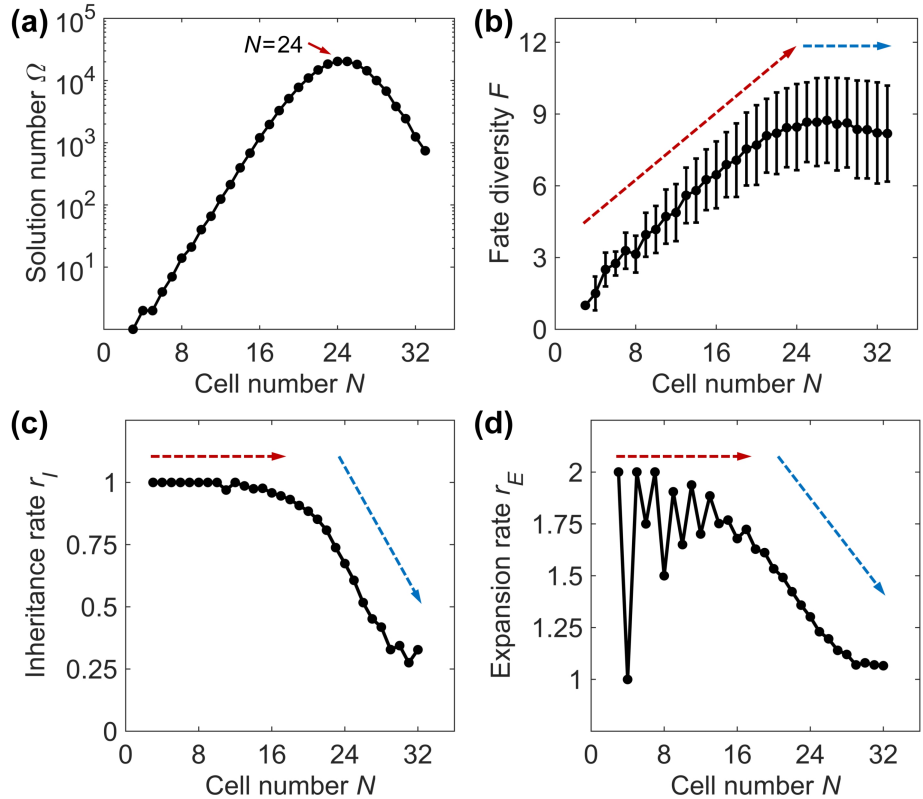

FIG. 5. Impact of target cell number on the solution space. (a) The number of distinct solutions. (b) Mean fate diversity against cell number. Error bar indicates the standard deviation of $F$ over lineage patterns at a given $N$. (c) The inheritance rate of robust solutions. (d) The expansion rate of distinct solutions.

*Discussion.* — Eutelic species has stereotypic developmental programs down to the single-cell level, especially for its cell lineage pattern. Why a nematode embryo programs cell divisions with specific order and segregation ratios has been elusive. In this work, a simple model is proposed to evaluate a lineage pattern's performance in proliferation speed, fate diversity and space robustness, on the basis of two assumptions derived from *C. elegans* early embryogenesis. The first is fate specification induced by asymmetric segregation of cell volume and its molecular contents; the second is the anti-correlation between cell cycle length and cell volume. Given that space robustness is essential for developmental accuracy at cellular resolution, we simulate and filter out lineage patterns with punctuated cell movement. The *C. elegans* pattern is well reproduced in our study with an outstanding proliferation speed, suggesting a minimum time principle at work. This principle was also found in development of intestinal crypts, which leads to a sharp transition between symmetric and asymmetric stem cell divisions [19]. The fitness gain from speed optimization may arise from competition among individuals, threat by predators and by internal environment hostile to cell or embryo [2].

Solutions (e.g., Patterns Nos. 3-8) that proliferate as fast as the *C. elegans* pattern might be used by other nematode species [20]. We identified a pattern which has one more asymmetric division compared to the *C. elegans* pattern in the first 4 cells, leading to significant asynchrony in their daughters, which is observed in a free-living marine nematode *Plectus sambesii* [21] [Pattern No. 9 in Fig. 3(a) (red diamond) and SM, Fig. S3]. Besides, other biological, biophysical or environmental factors may affect lineage pattern selection as well. For instance, the Pattern No. 5 has the same fate diversity to the *C. elegans* pattern, and even higher proliferation speed [SM, Fig. S3]. Why the *C. elegans* pattern is selected among those solutions remains to be answered. One possibility is that, activating asymmetric divisions only on stem cells involves the least amount of system resources such as genetic and proteomic programming [16]. Additionally, setting a relatively large volume segregation ratio during cytokinesis may be more robust for differentiation between the newborn sister cells.

This work provides a computational framework to simulate nematode lineage pattern, which can be further combined with developmental processes such as mechanical packing of cells and fate specification by cell-cell signaling [14, 18, 22-25]. The results provide new insights into the optimization principle in both embryogenesis and organogenesis [7, 26], and lay a foundation for rational design of functional multicellular systems [27, 28].

*Acknowledgements.* — We thank Feng Liu, Xiaojing Yang, Xiangyu Kuang and Kakit Kong for helpful discussions and comments. This work was supported by the National Natural Science Foundation of China (Grant Nos. 11635002, 12090053, 32088101, U1430237, U1530401), the Hong Kong Research Grants Council (Grants Nos. HKBU12100118, HKBU12100917, HKBU12123716, HKBU12324716, HKBU12301514) and the HKBU Interdisciplinary Research Cluster Fund. Computation was performed partly on the High Performance Computing Platform at Peking University.

[1] W. Tadroz and H. D. Lipshitz, Development **136**, 3033 (2009).

[2] P. H. O'Farrell, Cold Spring Harb. Perspect. Biol. **7**, a019042 (2015).

[3] P. Heyn, M. Kircher, A. Dahl, J. Kelso, P. Tomancak, A. T. Kalinka, and K. M. Neugebauer, Cell Rep. **6**, 285 (2014).

[4] M. T. Lee, A. R. Bonneau, and A. J. Giraldez, Annu. Rev. Cell Dev. Biol. **30**, 581 (2014).

[5] M. K. Wong, D. Guan, K. H. C. Ng, V. W. S. Ho, X. An, R. Li, X. Ren, and Z. Zhao, J. Biol. Chem. **291**, 12501 (2016).

[6] J. Nance and J. R. Priess, Development **129**, 387 (2002).

[7] J. E. Sulston, E. Schierenberg, J. G. White, and J. N. Thomson, Dev. Biol. **100**, 64 (1983).

[8] R. Schnabel and J. R. Priess, in *C. elegans II*, edited by D. R. Riddle et al. (Cold Spring Harbor Laboratory Press, 1997) pp. 361-382.

[9] J. Cao, G. Guan, M. K. Wong, L. Y. Chan, C. Tang, Z. Zhao, and H. Yan, bioRxiv (preprint) (2019), https://doi.org/10.1101/797688.

[10] G. Guan, M. K. Wong, V. W. S. Ho, X. An, L. Y. Chan, B. Tian, Z. Li, L.-H. Tang, Z. Zhao, and C. Tang, bioRxiv (preprint) (2019), https://doi.org/10.1101/776062.

[11] Z. Bao, Z. Zhao, T. J. Boyle, J. I. Murray, and R. H. Waterston, Dev. Biol. **318**, 65 (2008).

[12] L. Rose and P. Gönczy, in *WormBook*, edited by The C. elegans Research Community (2014), https://doi.org/10.1895/wormbook.1.30.2.

[13] Y. Arata, H. Takagi, Y. Sako, and H. Sawa, Front. Physiol. **5**, 529 (2015).

[14] R. Fickentscher, P. Struntz, and M. Weiss, Phys. Rev. Lett. **117**, 188101 (2016).

[15] R. Fickentscher, S. W. Krauss, and M. Weiss, New J. Phys. **20**, 113001 (2018).

[16] L. Hubatsch, F. Peglion, J. D. Reich, N. T. L. Rodrigues, N. Hirani, R. Illukkumbura, and N. W. Goehring, Nat. Phys. **15**, 1078 (2019).

[17] M. Laugsch and E. Schierenberg, Int. J. Dev. Biol. **48**, 655 (2004).

[18] B. Tian, G. Guan, L.-H. Tang, and C. Tang, Phys. Biol. **17**, 026001 (2020).

[19] S. Itzkovitz, I. C. Blat, T. Jacks, H. Clevers, and A. V. Oudenaarden, Cell **148**, 608 (2012).

[20] J. Schulze and E. Schierenberg, EvoDevo **2**, 18 (2011).

[21] J. Schulze, W. Houthoofd, J. Uenk, S. Vangestel, and E. Schierenberg, EvoDevo **3**, 13 (2012).

[22] R. Fickentscher, P. Struntz, and M. Weiss, Biophys. J. **105**, 1805 (2013).

[23] G. Guan, L.-H. Tang, and C. Tang, J. Phys.: Conf. Ser. **1592**, 012020 (2020).

[24] C. C. Mello, B. W. Draper, and J. R. Priess, Cell **77**, 95 (1994).

[25] C. J. Thorpe, A. Schlesinger, J. C. Carter, and B. Bowerman, Cell **90**, 695 (1997).

[26] J. E. Sulston, Philos. Trans. R. Soc. Lond. B **275**, 287 (1976).

[27] S. Toda, L. R. Blauch, S. K. Y. Tang, L. Morsut, and W. A. Lim, Science **361**, 156 (2018).

[28] S. Kriegman, D. Blackiston, M. Levin, and J. Bongard, Proc. Natl. Acad. Sci. U.S.A. **117**, 1853 (2020).